# Spatially resolved electronic inhomogeneities of graphene due to subsurface charges

*Andres Castellanos-Gomez[1,\*,+], Roel H. M. Smit[1,4], Nicolás Agraït[1,2,3] and Gabino Rubio-Bollinger[1,2].*

[1] Departamento de Física de la Materia Condensada (C–III). Universidad Autónoma de Madrid, Campus de Cantoblanco, E-28049 Madrid, Spain.

[2] Instituto Universitario de Ciencia de Materiales "Nicolás Cabrera". Universidad Autónoma de Madrid, Campus de Cantoblanco, E-28049 Madrid, Spain.

[3] Instituto Madrileño de Estudios Avanzados en Nanociencia IMDEA-Nanociencia, E-28049 Madrid, Spain.

[4] Kamerlingh Onnes Laboratory, Leiden University, Niels Bohrweg 2, 2333 CA Leiden, The Netherlands.

[+] Present address: Kavli Institute of Nanoscience, Delft University of Technology, Lorentzweg 1, 2628 CJ Delft (The Netherlands)

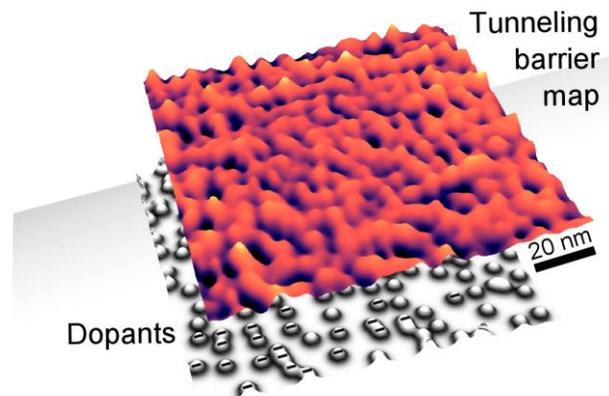

We probe the local inhomogeneities in the electronic properties of exfoliated graphene due to the presence of charged impurities in the $SiO_2$ substrate using a combined scanning tunneling and electrostatic force microscope. Contact potential difference measurements using electrostatic force microscopy permit us to obtain the average charge density but it does not provide enough resolution to identify individual charges. We find that the tunneling current decay constant, which is related to the local tunneling barrier height, enables one to probe the electronic properties of graphene distorted at the nanometer scale by individual charged impurities. We observe that such inhomogeneities do not show long range ordering and their surface density obtained by direct counting is consistent with the value obtained by macroscopic charge density measurements. These microscopic perturbations of the carrier density significantly alter the electronic properties of graphene, and their characterization is essential for improving the performance of graphene based devices.

---

[\*] Corresponding author. Tel/Fax: +31 (0) 152786272. E-mail address: a.castellanosgomez@tudelft.nl. (A. Castellanos-Gomez)

# 1. Introduction

The experimental realization of graphene by mechanical exfoliation of graphite on $SiO_2$ surfaces [1] has triggered a revolution in the design of electronic devices [2, 3] and chemical sensors [4, 5] because of the unique electronic properties [6-8] of graphene and its high sensitivity to the electrochemical environment [9]. Because of this high sensitivity, the electronic properties of graphene are strongly affected by substrate charged impurities [10, 11] leading to a large device-to-device variation in performance [11]. The charged impurities present in the substrate create an inhomogeneous electrostatic potential landscape [12, 13] with a typical length scale of tens of nanometers. These potential fluctuations cause a position dependent Dirac point on the graphene layer, producing spatial carrier density variations referred to in the literature as electron-hole puddles [14, 15].

Recently, efforts were made to fabricate suspended graphene devices [16, 17] in order to reduce an undesired interaction with the substrate. This strategy resulted in an increase in device reproducibility and superior electronic properties, such as a carrier mobility which can exceed $2 \cdot 10^5 \text{ cm}^2/\text{Vs}$. This improved performance shows the impact that substrate inhomogeneities have on the electronic properties of graphene [18]. Such a suspended geometry, however, increases the complexity of device fabrication. The development of alternative strategies and optimization of the fabrication of high performance and reproducible graphene based devices makes the study of the influence of charged impurities on the properties of graphene crucial. In recent publications Raman Spectroscopy [19] and Electrostatic Force Microscopy (EFM) [20] have been used to determine the density of charged impurities in graphene sheets. Although these experiments give a quantitative result, they do not provide a resolution sufficient to observe the effect of individual impurities on the properties of graphene. This has been achieved using a low temperature Scanning Tunneling Microscope (STM) to perform local tunneling spectroscopy [13]. Such a technique, however, cannot straightforwardly discriminate whether the local variation on the electronic properties of graphene is due to charged impurities or to graphene lattice defects [21] and demands a superior energy resolution requiring stringent cryogenic and ultrahigh vacuum conditions. This severely limits the use of this technique in the characterization of devices.

In this article we present a technique which enables us to determine the role of charged impurities on the spatial distribution of carrier density inhomogeneities even at room temperature and ambient conditions. An STM is used to obtain the topography of graphene and concomitantly a map of the local variations of the electron tunneling barrier height, which we observe to be very sensitive to spatial fluctuations of the carrier density. That is, the incompletely screened electric field produced by charged impurities strongly influences the band structure of graphene, affecting its tunneling barrier height and thus the tunneling current decay constant $\beta$, which we find to be spatially modulated at a scale close to 1 nm. From a statistical analysis of the spatial variations of $\beta$ in combination with EFM measurements one can determine the sign and the surface density of charged impurities which are the origin of the inhomogeneous properties of graphene.

This is the first time that the local changes of the tunnelling barrier height have been measured for an exfoliated graphene monolayer on top of $SiO_2$/Si surfaces. The tunneling barrier height is related to the work function and plays an important role in a variety of physical and chemical processes taking place on the graphene surface [22, 23]. Therefore the determination of the local changes of the tunnelling barrier height can be of crucial importance to characterize graphene-based electronic devices and chemical sensors. Additionally, this measurement allows sufficient resolution to spatially map the effect of a single charged impurity on the local electrical properties of the graphene layer. A better understanding of the nanoscale interaction between the charged impurities and a graphene layer will provide alternative ways to unleash the full potential of graphene in devices and sensors.

## 2. Sample preparation and characterization

Graphene samples were prepared by cleavage of highly oriented pyrolytic graphite (HOPG) using silicone stamps [24] instead of adhesive tape to minimize sample contamination. Regions where the graphene is only few atomic layers thick were identified, at first instance, by optical inspection. Then the optical contrast of these regions, measured at 9 different illumination wavelengths, has been fitted to a Fresnel law model to obtain the exact number of graphene layers [25, 26] (see supporting information). This optical characterization technique has been combined with atomic force microscopy measurements to reliably determine the thickness of the graphene layers. The electrical contact to the graphene flakes was provided by shadow-mask evaporation of a 30 nm thick gold layer as described in ref.[27].

In order to electrically characterize our graphene samples we used Electrostatic Force Microscopy (EFM) to sense the incompletely screened electric field caused by charged impurities measuring the contact potential difference ($V_{CPD}$) of the graphene layers [20]. The contact potential difference ($V_{CPD}$) measurements were carried out by placing the tip of our combined STM/AFM microscope about 20 nm above the surface of the graphene flake. We then applied a voltage ramp to the sample while measuring the resonance frequency shift of the force sensor, which is related to the electrostatic force gradient $\partial F/\partial z$ [28]. The force gradient ($\partial F/\partial z$) has a parabolic dependence with the tip-sample voltage and its vertex is at the voltage that counteracts $V_{CPD}$ (for more detailed description see supplementary information).

For increasing thickness of graphene flakes the electric field generated by the charged impurities is increasingly screened and the $V_{CPD}$ approaches the bulk value [20, 29]. For small thicknesses, the sign and magnitude of the deviation from the bulk value of the $V_{CPD}$ ($\Delta V_{CPD}$) is related to the sign and the density of the charged impurities [20]. The $\Delta V_{CPD}$ decreases with flake thickness, as shown in Figure 1, indicating the presence of negatively charged impurities in the substrate. The maximum deviation of $\Delta V_{CPD}$ is found for the monolayer graphene being $\Delta V_{CPD} = -0.46 \pm 0.03$ V from which one can estimate [20] a surface density of negatively charged impurities on the order of $10^{12}$ cm$^{-2}$. The presence of negative charged impurities in the SiO$_2$ substrate is common in graphene field-effect transistor devices fabricated in air [18], showing a marked p-type behavior, and has been previously attributed to an interfacial defect or impurity layer [20]. Up to now, EFM experiments on exfoliated graphene have shown lateral resolution in the order of 50-100 nm which is not enough to resolve the effect of an individual charged impurity on the electronic properties of graphene.

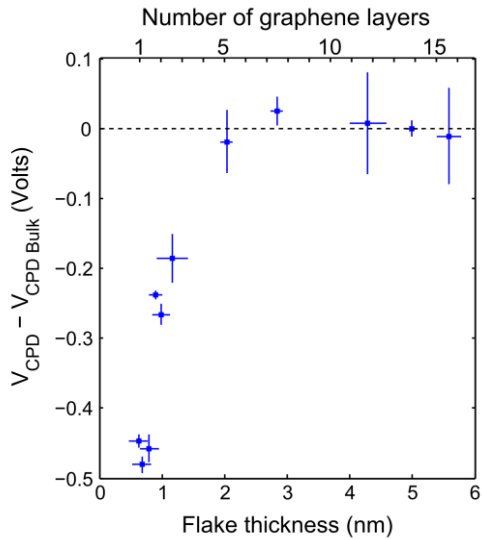

**Figure 1.** Dependence of the deviation from the bulk value of the $V_{CPD}$ as a function of the flake thickness. This thickness dependence of the $V_{CPD}$ is caused by the electric field originated by charged impurities in the substrate which is incompletely screened by thin flakes. The $V_{CPD}$ measured in thin flakes is lower than the bulk value indicating the presence of negatively charged impurities in the substrate.

## 3. Results and discussion

Figure 2 shows the topography of a mechanically exfoliated graphite flake deposited on a 285 nm SiO$_2$/Si substrate obtained in the constant current STM mode. All STM measurements were acquired under ambient conditions with a homebuilt combined STM/AFM [30-32] microscope supplemented with a *qPlus* force sensor [33] (spring constant $k \sim 12500$ N/m, resonance frequency $f_0 \sim 32.1$ kHz and quality factor $Q \sim 4200$) using carbon fiber tips [34]. The topography shows a region where diverse graphite thicknesses are found down to a single layer of graphene. One clearly notices that the thinner the layer the stronger the corrugation. The roughness average of the height coordinate $z$, given by $R_a = \frac{1}{N}\sum_{j=1}^{N}|z_j - \bar{z}|$ (where $N$ is the total number of pixels), is $R_a = 0.13 \pm 0.03$ nm for the monolayer region and $R_a = 0.06 \pm 0.02$ nm for the multilayered regions. Previous studies have shown that the corrugation of graphene includes two components [35]: one originating from the roughness of the SiO$_2$ substrate [35, 36] and another from the intrinsic rippling of the graphene sheet [35, 37]. Constant current STM images, such as Figure 2, reliably represent the surface topography only for samples with homogeneous electronic properties. Before concluding that this corrugation is purely structural one, thus, has to exclude spatial variations in its electronic properties. While metallic samples in general satisfy this demand, this is not straightforward for an atomically thin layer of graphene, since even for the cleanest of sample fabrication techniques [38] one cannot rule out the presence of charged impurities close to the graphene layer [13]. Such impurities will induce a local carrier doping which modifies the electronic properties of the graphene layer at the nanometer scale [13, 39].

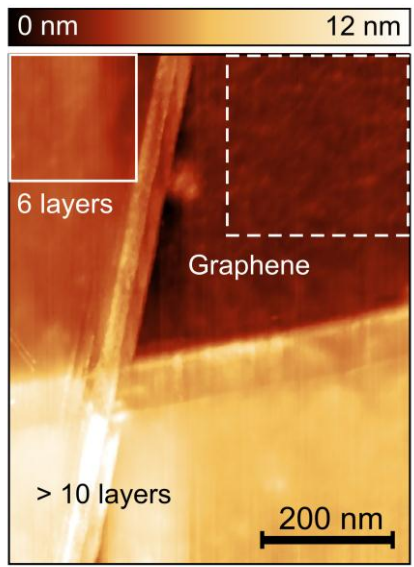

**Figure 2.** Constant current STM topography (590 nm x 750 nm) of an exfoliated graphite flake deposited on a 285 nm SiO$_2$/Si substrate. The STM parameters are tunneling current $I = 0.8$ nA and bias voltage $V_{sample} = 0.1$ V. The number of graphene layers of the different regions in the flake has been determined by the combination of AFM measurements with quantitative optical microscopy. The measured roughness is significantly larger in the monolayer than in the multilayer regions.

To probe the presence of these impurities and their impact on the electronic properties of graphene an additional and independent measurement is required. It has been shown that the presence of charges on a substrate can be detected by measuring the local tip-sample interaction [40, 41]. The AFM capability of our combined STM/AFM has been used to measure the tip-sample interaction, simultaneously with the STM topography, using the frequency modulation AFM mode [42]. The topography and the tip-sample interaction simultaneously measured on top of the single graphene layer can be seen in Figure 3a and 3b respectively. Charged impurities in the substrate, which cause an attractive electrostatic interaction between tip and sample, appear as dark dips in Figure 2b. When comparing both images in Figure 3 it can be clearly observed that a bright hillock in the apparent topography of the sample is frequently accompanied by such a dip in the tip-sample interaction (see the arrows in Figures 3a and 3b). To quantify this similarity between the two images we calculate their normalized cross correlation (Figure 3c) which shows a minimum at the origin with a value of -0.7 with a Full Width at Half Minimum (FWHM) of 10 nm. The presence of such a narrow and prominent negative peak at the origin of this plot implies a strong anti-correlation between Figure 3a and 3b. The anti-correlation between the topography by STM and the tip-

- 5 -

sample electrostatic interaction indicates that there is a contribution to the apparent corrugation which is due to the inhomogeneous electronic properties of graphene induced by the presence of subsurface charged impurities. We find that this contribution can be the dominant source of the apparent corrugation which can be additionally influenced by the substrate roughness [35, 36] and the intrinsic rippling of graphene [35, 37].

On the other hand, if we simultaneously measure the tip-sample interaction (Figure 4b) with the topography (Figure 4a) on top of a multilayer region, we do not observe these strong localized inhomogeneities in the tip-sample interaction image. Indeed, the calculated cross-correlation between the topography and tip-sample interaction images (Figure 4c) shows less pronounced anti-correlation than for the single graphene layer, see Figure 3c. There is only an ill-defined peak, slightly displaced from the origin, with a value of -0.4 and a FWHM of 26 nm. Because of this weak anti-correlation one can conclude that the topographic corrugation on the multilayer region is less influenced by the presence of the charged impurities as they are more effectively screened which is in agreement with the EFM measurements shown in Figure 1.

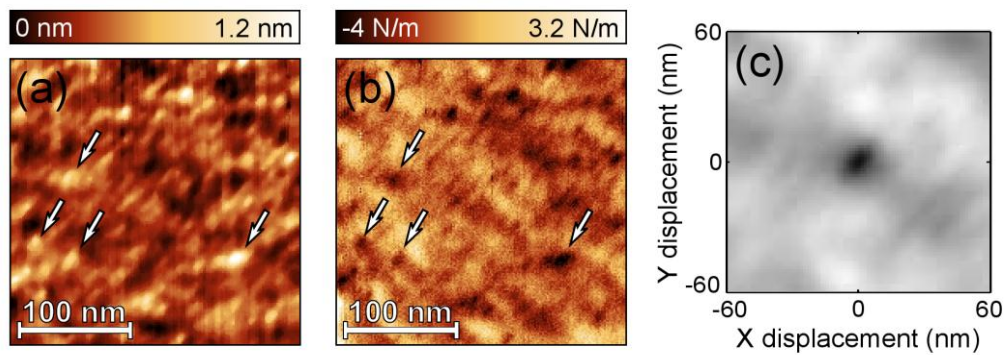

**Figure 3.** (a) Constant current STM topography of the single layer region marked by the dashed square (270 nm x 270 nm) in Figure 1. The image has been obtained in dynamic STM mode with time averaged tunneling current $\langle I \rangle = 0.8$ nA, bias voltage $V_{sample} = 0.1$ V and tip oscillation amplitude $A_{osc} = 0.2$ nm$_{RMS}$. (b) Tip-sample electrostatic interaction simultaneously measured during the STM topography scan. The presence of charged impurities in the substrate locally causes an attractive tip-sample electrostatic interaction which can be identified as well-defined dark spots in the tip-sample interaction image. The arrows mark some bright hillocks in (a) and their corresponding dips in (b), indicating a high degree of anti-correlation between these two images. (c) Calculated normalized cross-correlation between the STM topography and the tip-sample interaction used to quantify the relationship between them. The presence a marked negative peak at the origin of (c) proves the anti-correlation between the topography and the tip-sample interaction images and indicates that the STM topography of graphene is affected by the presence of subsurface charges.

- 6 -

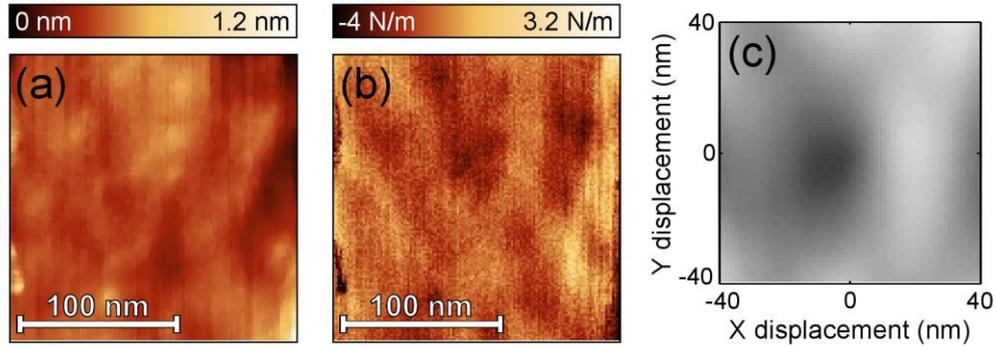

**Figure 4.** (a) Constant current STM topography of the multilayer graphene region marked by the solid square (190 nm x 190 nm) in Figure 1 obtained in dynamic STM mode. The dynamic STM parameters are time averaged tunneling current $\langle I \rangle = 0.8$ nA, bias voltage $V_{sample} = 0.1$ V and tip oscillation amplitude $A_{osc} = 0.2$ nm$_{RMS}$. (b) Tip-sample electrostatic interaction measured simultaneously during the STM topography scan. The absence of marked inhomogeneities in the tip-sample image indicates that the electric field originated by the subsurface charges is completely screened. (c) Calculated normalized cross-correlation between the STM topography and the tip-sample interaction which shows a less pronounced anti-correlation than for the single graphene layer.

Because the electrostatic forces acting between tip and sample are long range the lateral resolution is insufficient to individually resolve the charges and their effect on graphene. We therefore employ an STM-based technique [43] which benefits from the exponential dependence of the tunneling current ($I \propto e^{-\beta z}$) upon the tip-sample distance (*z*). It therefore constitutes a convenient probe to measure the local variations of the electronic properties of graphene induced by charged impurities.

We observe that the tunneling current decay constant ($\beta$) is strongly influenced by the presence of subsurface charges [44, 45] and the lateral resolution is sufficient to identify individual carrier inhomogeneities in graphene. The physics that is at the basis of this can be understood as follows: the presence of negatively (positively) charged impurities causes an electric field which effectively shifts the energy of the bottom of the band by an amount $\Delta E$. This effect reduces (increases) the apparent tunneling barrier height, $\Phi_{app}$ (Figure 5c). This barrier height change can be modeled as $\Phi_{app} = (\Phi_{graphene} - \Phi_{tip} \mp \Delta E)/2$ where $\Phi_{graphene}$ and $\Phi_{tip}$ are the graphene and the tip work function respectively. The associated tunneling decay constant then changes according to $\beta = 2\sqrt{2m \cdot \Phi_{app}}/\hbar$ (*m* is the electron mass and $\hbar$ is the reduced Planck constant). To measure $\beta$ we use the dynamic STM operation mode [46] in which the tip-sample distance ($\Delta z$) is modulated at sub-nanometer oscillation amplitude and the tunneling current oscillation amplitude ($\Delta I$) is measured using a homemade current-to-voltage converter with a bandwidth > 30 kHz. From this, one can obtain $\beta = -(\Delta I/\Delta z)/\langle I \rangle$ with $\langle I \rangle$ the time averaged tunneling current. While the measurement of $\beta$ can also be done by the standard operation mode in which one records a map of current versus distance traces at every point of the image, we opt for this dynamic STM scheme which allows to scan at regular speed, and facilitates the simultaneous measurement of the interaction force by dynamic AFM. The dynamic STM measurements were carried out by keeping the time averaged tunneling current constant with a feedback control loop while the tip oscillates at the resonance frequency of the *qPlus* sensor. The tip oscillation amplitude was 0.2 nm$_{pp}$. Variations of the resonance frequency of the *qPlus* sensor were measured with the phase locked loop of a *Nanonis OC4 Oscillation Controller*. The tunneling current decay constant, *β*, has been obtained from the derivative of the tunneling current against the tip-sample distance ($\beta = (\partial I/\partial z)/I$), measured with a lock in amplifier *SR830 Stanford Research Systems* (schematic diagrams of the experimental setup can be found in the supplementary information). In order to calibrate the tunneling current decay constant, measured by dynamic STM, we have employed quasi-static current *vs*. distance traces measured in a multilayer region. In this region the tunneling decay constant measured by both the quasi-static current *vs*. distance experiments and dynamic STM shows a well-defined value, providing the calibration (See supporting information).

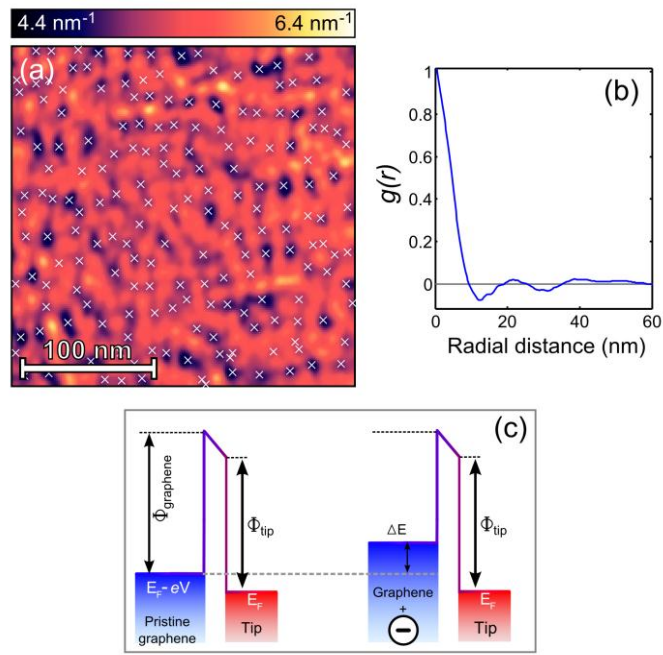

**Figure 5.** (a) Spatial variation of the tunneling current decay constant $\beta$ on graphene (over an area of 270 nm x 270 nm) simultaneously measured with the STM topography and the tip-sample electrostatic interaction shown in Figures 2a and 2b respectively. The image has been obtained in dynamic STM mode ($\langle I \rangle = 0.8$ nA, $V_{sample} = 0.1$ V and $A_{osc} = 0.2$ nm$_{RMS}$). The $\beta$ image shows strong localized inhomogeneities (marked with white crosses) caused by the local modification of the tunneling barrier due to the presence of individual subsurface charges. (b) Radially averaged two dimensional autocorrelation function $g(r)$ of the tunneling decay constant $\beta$ image in (a) used to statistically analyze the distribution of localized inhomogeneities. The obtained average spacing between dips is 20 nm and their average radius is 5 nm. The lack of a well-defined periodic oscillation of $g(r)$ indicates the absence of long-range ordering in the distribution. (c) Simplified one dimensional tunneling diagram which illustrates the modification of the tunneling barrier caused by the electric field originated by a negative charged impurity in the substrate. The band structure of graphene shifts upwards, effectively reducing the apparent tunneling barrier height and thus the tunneling decay constant.

Figure 5a shows the spatial variation of $\beta$ for a single graphene layer simultaneously measured with the STM topography (Figure 2a) and the tip-sample electrostatic interaction (Figure 2b). The $\beta$ image shows a low average value ($5.2 \pm 0.4$ nm$^{-1}$), typical of STM operation in air [47], and localized inhomogeneities (identified as dips in the $\beta$ image) caused by local doping induced by negative charged impurities (Figure 5c). Considering that each dip in the $\beta$ image is due to the presence of one individual negative charge we can estimate the density of charged impurities by tentatively counting the number of depressions in a given area, resulting in an impurity density $\sigma = (2.9 \pm 0.6) \cdot 10^{11}$ cm$^{-2}$. A more objective procedure to statistically analyze the short-range ordering [48] and spatial variation of $\beta$ is to obtain the radially-averaged autocorrelation function $g(r)$, shown in Figure 5c. The typical radius $r_{FWHM}$ at half minimum of the localized inhomogeneities is obtained from the radial distance at which the value of $g(r)$ is 0.5, which yields $r_{FWHM} = 4.7 \pm 1.1$ nm. The inhomogeneities do not show long range ordering and the mean spacing between them is $d = 22 \pm 2$ nm, determined from the position of the first maximum of $g(r)$ [48], which corresponds to a charge density $\sigma = \pi^{-1}(d/2)^{-2} = (2.6 \pm 0.5) \cdot 10^{11}$ cm$^{-2}$. When this measurement is carried out in the multilayer region one obtains a nearly constant value of $\beta$ without spatial inhomogeneities (see Supplementary information) which indicates that the electric field induced by the charged impurities is screened by the layers. These results are agreement with the EFM measurement shown in Figure 1 and with the combined STM/AFM measurements shown in Figures 3 and 4.

We have also check the cross-correlation between the apparent roughness of the graphene surface, measured in constant current STM mode and the presence of the inhomogeneities in the $\beta$ maps, finding that the $\beta$ maps in graphene are anti-correlated with the STM topography (see Supplementary Information). Therefore the observed $\beta$ inhomogeneities induced by charged impurities are responsible for an apparent STM-measured topographic roughness, as that shown in Figure 3a, unrelated to substrate roughness.

## 4. Summary

We have studied the effect of subsurface charged impurities on the local electronic properties of graphene deposited by mechanical exfoliation on silicon oxide. We have employed a combined STM/AFM which allows for simultaneous measurement of sample topography, tip-sample electrostatic forces and tunneling current decay constant and to study their correlation. The sign and density of the substrate charged impurities have been obtained from the measurement of the contact potential difference of graphene using electrostatic force microscopy, but this operation mode does not provide enough lateral resolution to probe the effect of an individual charged impurity on the electronic properties of graphene. We have found that the tunneling current exponential decay constant, related to the local tunneling barrier height, is very sensitive to the presence of subsurface charges and is spatially inhomogeneous. Consequently the apparent topography measured in constant current STM images is not only related to the roughness of graphene but also dependent on the spatial distortions of the electronic structure of graphene. The local tunneling decay constant image shows a distribution of localized inhomogeneities which are related to the presence of negative subsurface charges. In the studied samples these inhomogeneities do not show long-range ordering and their average spacing is 22 nm with an average radius of 5 nm. The average charge density obtained from the electrostatic force microscopy measurements is consistent with the charge density that can be obtained by counting the number of spots in the graphene layer where its local tunneling barrier height is strongly distorted. Such local tunneling decay constant measurements are robust making them attractive for high resolution microscopic characterization of graphene devices.

**Acknowledgements**

A.C-G. acknowledges fellowship support from the Comunidad de Madrid (Spain). This work was supported by MICINN (Spain) GR/MAT/0111/2004, MAT2008-01735 and CONSOLIDER-INGENIO-2010 'Nanociencia Molecular' CSD-2007-00010.

**References**

[1] Novoselov KS, Geim AK, Morozov SV, Jiang D, Zhang Y, Dubonos SV, et al. Electric field effect in atomically thin carbon films. Science. 2004;306(5696):666-9.
[2] Lin Y-M, Jenkins KA, Valdes-Garcia A, Small JP, Farmer DB, Avouris P. Operation of Graphene Transistors at Gigahertz Frequencies. Nano Lett. 2008;9(1):422-6.
[3] Xia F, Mueller T, Lin Y, Valdes-Garcia A, Avouris P. Ultrafast graphene photodetector. Nature Nanotech. 2009;4(12):839-43.
[4] Dan Y, Lu Y, Kybert NJ, Luo Z, Johnson ATC. Intrinsic response of graphene vapor sensors. Nano Lett. 2009;9(4):1472-5.
[5] Wehling TO, Novoselov KS, Morozov SV, Vdovin EE, Katsnelson MI, Geim AK, et al. Molecular doping of graphene. Nano Lett. 2008;8(1):173-7.
[6] Novoselov KS, Jiang Z, Zhang Y, Morozov SV, Stormer HL, Zeitler U, et al. Room-temperature quantum Hall effect in graphene. Science. 2007;315(5817):1379.
[7] Ozyilmaz B, Jarillo-Herrero P, Efetov D, Abanin DA, Levitov LS, Kim P. Electronic transport and quantum hall effect in bipolar graphene p-n-p junctions. Phys Rev Lett. 2007;99(16):166804.
[8] Zhang Y, Tan YW, Stormer HL, Kim P. Experimental observation of the quantum Hall effect and Berry's phase in graphene. Nature. 2005;438(7065):201-4.
[9] Chen F, Qing Q, Xia J, Li J, Tao N. Electrochemical Gate-Controlled Charge Transport in Graphene in Ionic Liquid and Aqueous Solution. J Am Chem Soc. 2009;131(29):9908-9.
[10] Chen JH, Jang C, Adam S, Fuhrer MS, Williams ED, Ishigami M. Charged-impurity scattering in graphene. Nature Phys. 2008;4(5):377-81.

[11]   Tan YW, Zhang Y, Bolotin K, Zhao Y, Adam S, Hwang EH, et al. Measurement of Scattering Rate and Minimum Conductivity in Graphene. Phys Rev Lett. 2007;99(24):246803.

[12]   Adam S, Hwang EH, Rossi E, Das Sarma S. Theory of charged impurity scattering in two-dimensional graphene. Solid State Commun. 2009;149(27-28):1072-9.

[13]   Zhang Y, Brar VW, Girit C, Zettl A, Crommie MF. Origin of spatial charge inhomogeneity in graphene. Nature Phys. 2009;5(10):722-6.

[14]   Martin J, Akerman N, Ulbricht G, Lohmann T, Smet JH, Von Klitzing K, et al. Observation of electron–hole puddles in graphene using a scanning single-electron transistor. Nature Phys. 2008;4(2):144-8.

[15]   Deshpande A, Bao W, Miao F, Lau CN, LeRoy BJ. Spatially resolved spectroscopy of monolayer graphene on SiO2. Phys Rev B. 2009;79(20):205411.

[16]   Bolotin KI, Sikes KJ, Jiang Z, Klima M, Fudenberg G, Hone J, et al. Ultrahigh electron mobility in suspended graphene. Solid State Commun. 2008;146(9-10):351-5.

[17]   Du X, Skachko I, Barker A, Andrei EY. Approaching ballistic transport in suspended graphene. Nature Nanotech. 2008;3(8):491-5.

[18]   Romero HE, Shen N, Joshi P, Gutierrez HR, Tadigadapa SA, Sofo JO, et al. n-Type Behavior of Graphene Supported on Si/SiO2 Substrates. ACS nano. 2008;2(10):2037-44.

[19]   Ni ZH, Yu T, Luo ZQ, Wang YY, Liu L, Wong CP, et al. Probing charged impurities in suspended graphene using Raman spectroscopy. ACS nano. 2009;3(3):569-74.

[20]   Datta SS, Strachan DR, Mele EJ, Johnson ATC. Surface Potentials and Layer Charge Distributions in Few-Layer Graphene Films. Nano Lett. 2008;9(1):7-11.

[21]   Rutter GM, Crain JN, Guisinger NP, Li T, First PN, Stroscio JA. Scattering and Interference in Epitaxial Graphene. Science. 2007;317(5835):219-22.

[22]   Shi Y, Kim KK, Reina A, Hofmann M, Li LJ, Kong J. Work function engineering of graphene electrode via chemical doping. ACS nano. 2010;4(5):2689-94.

[23]   Yi Y, Choi WM, Kim YH, Kim JW, Kang SJ. Effective work function lowering of multilayer graphene films by subnanometer thick AlO overlayers. Appl Phys Lett. 2011;98:013505.

[24]   Moreno-Moreno M, Castellanos-Gomez A, Rubio-Bollinger G, Gomez-Herrero J, Agraït N. Ultralong Natural Graphene Nanoribbons and Their Electrical Conductivity. Small. 2009;5(8):924-7.

[25]   Blake P, Hill EW, Neto AHC, Novoselov KS, Jiang D, Yang R, et al. Making graphene visible. Appl Phys Lett. 2007;91(6):063124.

[26]   Castellanos-Gomez A, Agrait N, Rubio-Bollinger G. Optical identification of atomically thin dichalcogenide crystals. Appl Phys Lett. 2010;96(21):213116-3.

[27]   Staley N, Wang H, Puls C, Forster J, Jackson TN, McCarthy K, et al. Lithography-free fabrication of graphene devices. Appl Phys Lett. 2007;90:143518.

[28]   Giessibl F. A direct method to calculate tip–sample forces from frequency shifts in frequency-modulation atomic force microscopy. Appl Phys Lett. 2001;78:123.

[29]     Yu YJ, Zhao Y, Ryu S, Brus LE, Kim KS, Kim P. Tuning the graphene work function by electric field effect. Nano Lett. 2009;9(10):3430-4.

[30]     Smit R, Grande R, Lasanta B, Riquelme J, Rubio-Bollinger G, Agraït N. A low temperature scanning tunneling microscope for electronic and force spectroscopy. Rev Sci Instrum. 2007;78:113705.

[31]     Castellanos-Gomez A, Agrait N, Rubio-Bollinger G. Dynamics of quartz tuning fork force sensors used in scanning probe microscopy. Nanotechnology. 2009;20(21):215502.

[32]     Castellanos-Gomez A, Agrait N, Rubio-Bollinger G. Force-gradient-induced mechanical dissipation of quartz tuning fork force sensors used in atomic force microscopy. Ultramicroscopy. 2011;111(3):186-90.

[33]     Giessibl FJ. High-speed force sensor for force microscopy and profilometry utilizing a quartz tuning fork. Appl Phys Lett. 1998;73(26):3956-8.

[34]     Castellanos-Gomez A, Agrait N, Rubio-Bollinger G. Carbon fibre tips for scanning probe microscopy based on quartz tuning fork force sensors. Nanotechnology. 2010;21(14):145702.

[35]     Geringer V, Liebmann M, Echtermeyer T, Runte S, Schmidt M, Rückamp R, et al. Intrinsic and extrinsic corrugation of monolayer graphene deposited on SiO2. Phys Rev Lett. 2009;102(7):076102.

[36]     Ishigami M, Chen J, Cullen W, Fuhrer M, Williams E. Atomic structure of graphene on SiO 2. Nano Lett. 2007;7(6):1643-8.

[37]     Meyer JC, Geim AK, Katsnelson MI, Novoselov KS, Booth TJ, Roth S. The structure of suspended graphene sheets. Nature. 2007;446(7131):60-3.

[38]     Bao W, Liu G, Zhao Z, Zhang H, Yan D, Deshpande A, et al. Lithography-free fabrication of high quality substrate-supported and freestanding graphene devices. Nano Research. 2010;3(2):98-102.

[39]     Katsnelson MI, Guinea F, Geim AK. Scattering of electrons in graphene by clusters of impurities. Phys Rev B. 2009;79(19):195426.

[40]     Seo Y, Jhe W, Hwang C. Electrostatic force microscopy using a quartz tuning fork. Appl Phys Lett. 2002;80:4324.

[41]     Gross L, Mohn F, Liljeroth P, Repp J, Giessibl FJ, Meyer G. Measuring the Charge State of an Adatom with Noncontact Atomic Force Microscopy. Science. 2009;324(5933):1428-31.

[42]     Albrecht T, Grütter P, Horne D, Rugar D. Frequency modulation detection using high-Q cantilevers for enhanced force microscope sensitivity. J Appl Phys. 1991;69(2):668.

[43]     de Kort R, van der Wielen MCMM, van Roij AJA, Kets W, van Kempen H. Zn- and Cd-induced features at the GaAs(110) and InP(110) surfaces studied by low-temperature scanning tunneling microscopy. Phys Rev B. 2001;63(12):125336.

[44]     Kobayashi K, Kurokawa S, Sakai A. Scanning tunneling microscopy and barrier-height imaging of subsurface dopants on GaAs (110). Jpn J Appl Phys. 2005;44(12):8619.

[45]     Kurokawa S, Takei T, Sakai A. A search for subsurface dopants on hydrogen-terminated Si (111) surfaces. Jpn J Appl Phys. 2003;1(42):4655.

[46]   Herz M, Schiller C, Giessibl FJ, Mannhart J. Simultaneous current-, force-, and work-function measurement with atomic resolution. Appl Phys Lett. 2005;86(15).

[47]   Meepagala S, Real F. Detailed experimental investigation of the barrier-height lowering and the tip-sample force gradient during STM operation in air. Phys Rev B. 1994;49(15):10761-3.

[48]   Sánchez F, Infante IC, Lüders U, Abad L, Fontcuberta J. Surface roughening by anisotropic adatom kinetics in epitaxial growth of La0. 67Ca0. 33MnO3. Surf Sci. 2006;600(6):1231-9.

## Supporting information:

# Spatially resolved electronic inhomogeneities of graphene due to subsurface charges

*Andres Castellanos-Gomez[1,+], Roel H. M. Smit[1,4], Nicolás Agraït[1,2,3] and Gabino Rubio-Bollinger[1,2]*.

[1] Departamento de Física de la Materia Condensada (C–III). Universidad Autónoma de Madrid, Campus de Cantoblanco, E-28049 Madrid, Spain.

[2] Instituto Universitario de Ciencia de Materiales "Nicolás Cabrera". Universidad Autónoma de Madrid, Campus de Cantoblanco, E-28049 Madrid, Spain.

[3] Instituto Madrileño de Estudios Avanzados en Nanociencia IMDEA-Nanociencia, E-28049 Madrid, Spain.

[4] Kamerlingh Onnes Laboratory, Leiden University, Niels Bohrweg 2, 2333 CA Leiden, The Netherlands.

[+]Present address: Kavli Institute of Nanoscience, Delft University of Technology, Lorentzweg 1, 2628 CJ Delft (The Netherlands)

E-mail: a.castellanosgomez@tudelft.nl

### *Electrodes deposition*

As graphene consists of a single atomic layer it is particularly sensitive to surface contaminants. It has been reported how the contamination produced by the lithography processes can cover up to a 30% of the graphene surface[1] and in some cases it can make impossible to achieve atomic resolution STM images.[2] This fact has motivated the development of annealing techniques[2-4] to clean the graphene surface after sample fabrication. Another strategy to avoid the contamination of the graphene during the lithography is using a lithography-free technique to deposit electrodes on the graphene samples. Different procedures have been developed in the last years such as shadow mask evaporation,[5] stencil mask evaporation[6] and direct microsoldering.[7] We have used the shadow mask evaporation technique described in ref.[5].

A 7 µm in diameter carbon fiber, obtained from a carbon fiber rope, was placed onto the flake of interest using a 3-axis micropositioner. After the fiber is located in the desired region, another carbon fiber can be placed over the first one forming a cross-shaped shadow mask. Figure S1a shows an example of a few layers graphene flake covered by a shadow mask formed by two crossed carbon fibers. After that, a 30 nm thick gold film has been thermally evaporated creating 4 electrodes which are shown in Figure S1b. We have found this cross-shaped geometry very useful for the positioning of the STM tip at the desired region.

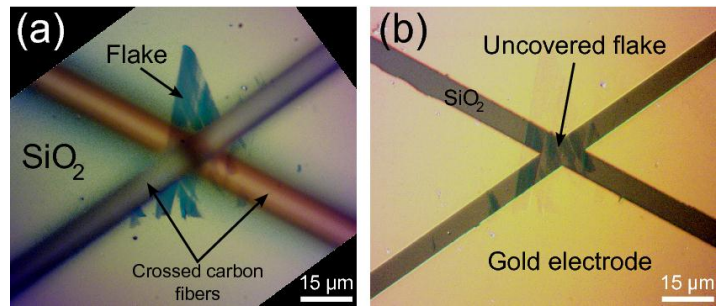

**Figure S1:** (a) Few layers graphene flake deposited on 285 nm $SiO_2$ layer. Two crossed carbon fibers cover part of the flake to form the shadow mask. (b) Gold electrodes deposited by thermal evaporation of 30 nm of gold.

## *Sample characterization*

In order to identify the number of graphene layers in a fast and non-destructive way we have quantitatively analyzed optical microscopy images of the samples. First, fabricated samples have been inspected under a Nikon Eclipse LV-100 optical microscope using a 50× objective with 0.8 numeric aperture to locate the thinnest flakes. Then the optical contrast of these flakes has been measured[8] using an EO-1918C 1/1.8" camera from Edmund Optics and nine narrow bandpass filters to select specific illumination wavelengths in the visible spectrum. The thickness of these flakes has been obtained from a fit of the optical contrast experimental values to a Fresnel law based model[9] in where the only free parameter is the number of graphene monolayers. Figure S2 shows the measured optical contrast for three flakes with different thicknesses. A careful comparison between the thickness obtained by this procedure and the one measured by means of atomic force microscopy reveals that the thickness measured by AFM is systematically 0.4 nm larger which is compatible with the presence of a layer attributed[10] to adsorbed water under the flake.

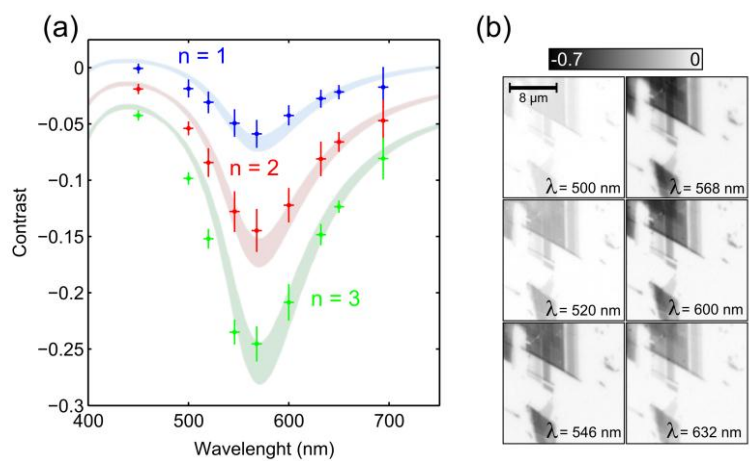

**Figure S2:** (a) Optical contrast *vs.* illumination wavelength measured for three flakes with different thicknesses. From the fit to the Fresnel law model (solid lines) we estimate their number of graphene layers *n*. The traces corresponding to n=1,2,3 have been vertically displaced for clarity by 0,-0.025 and -0.05 respectively. (b) Optical contrast maps measured for 6 different illumination wavelengths, all sharing the same contrast bar for easier comparison.

## *Contact potential difference measurement*

To measure the contact potential difference we first place the STM/AFM tip about 20 nm above the surface of the graphene flake. We then apply a voltage ramp to the sample while measuring the resonance frequency shift of the force sensor, which is related to the electrostatic force gradient $\partial F/\partial z$ by

$$\frac{\partial F}{\partial z} = -\frac{1}{2}\frac{\partial^2 C(z)}{\partial z^2}\cdot(V_{DC}-V_{CPD})^2,$$

where $C$ is the tip-sample capacitance, $V_{DC}$ is the tip-sample bias voltage and $V_{CPD}$ is the contact potential difference due to the different work function ($\Phi$) of tip and sample.

The force gradient ($\partial F/\partial z$) has a parabolic dependence with the tip-sample voltage and its vertex is at voltage that counteracts $V_{CPD}$. Thus, placing the tip on areas with different thickness and measuring the parabolic dependence of the force gradient with the tip-sample voltage one can obtain the relationship between the $V_{CPD}$ and the number of graphene layers (see Figure S3).

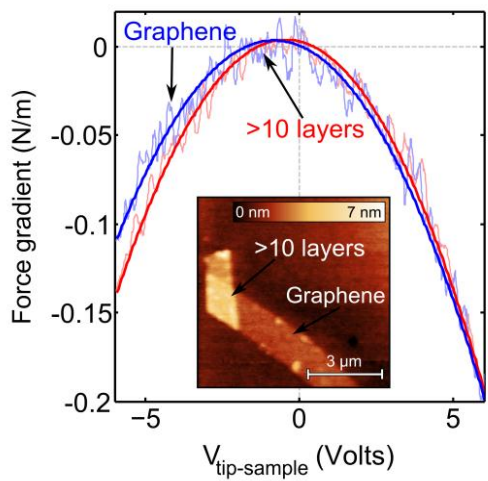

**Figure S3:** Electrostatic force gradient as a function of the applied tip-sample bias voltage, measured in two regions with different number of graphene layers. In the monolayer graphene region, the voltage of the parabola vertex is lower than in the multilayer region. We conclude, therefore, that in our sample the contact potential difference measured for graphene is lower than for the multilayer. (Insert) Topography in FM-AFM mode of the studied graphene flake.

## *STM/AFM measurements on graphene flakes*

By using a combined STM/AFM we can study the electronic properties of conductive nanopatches eventhough they are deposited on top of insulating surfaces. First, the carbon fiber tip is located on top of the desired graphene flake with the help of a long working distance optical microcope.[11] After that, we make use of the AFM capabilities of our combined STM/AFM to scan the region under study. The AFM measurements have been carried out in the frequency modulation mode (FM-AFM) using the oscillation amplitude as feedback signal (see Figure S4) to keep the tip sample distance constant and obtain the sample topography (see Figure S5a). In this way the changes in the resonance frequency, related to changes in the tip-sample interaction, can be simultaneously measured by means of a *phase* *locked loop* PLL (Figure S5b). This change in the resonance frequency has been used to distinguish the SiO$_2$ from the flake regions because of their different tip-sample interaction.

Once the flake region is identified, more detailed AFM topography images are used to determine the thickness of the different areas in the flake as shown in Figure S5c-d. Before the combined STM/AFM measurement starts the tip is positioned onto the flake, which is electrically contacted by a gold electrode, and the scan range is reduced (dashed square in Figure S5c) in order to avoid the tip reaching the insulator substrate which would result in a tip crash that degrades the sample and/or tip irreversibly.

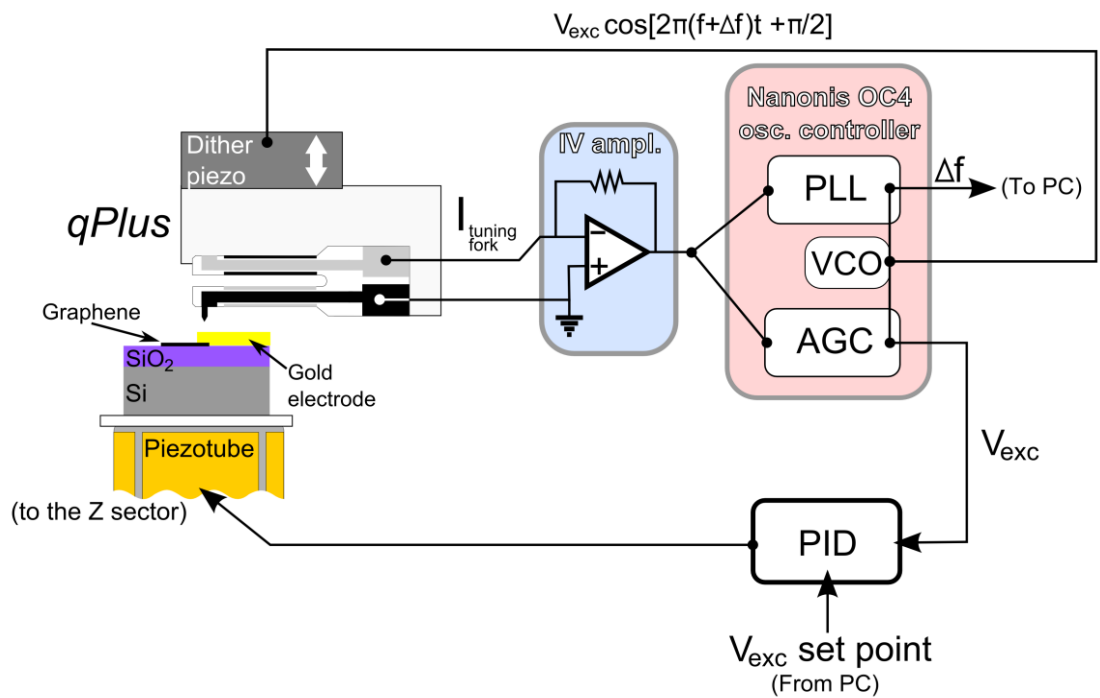

**Figure S4:** Schematic diagram of the experimental setup used in the AFM measurements of the topography of graphene crystals. The tuning fork based force sensor, in *qPlus* configuration, is mechanically excited by a dither piezoelectric actuator. The piezoelectric current generated by the oscillation of the (free) prong is preamplified by a current-to-voltage converter (gain × 10$^8$). The converter is connected to a *Nanonis OC4 Oscillation Controller* with a phase locked loop (PLL) and an automatic gain control (AGC). The output of *OC4 Nanonis* provides an excitation signal, which is connected to the dither piezo, at the resonance frequency of the sensor ($f + \Delta f$) and phase shifted 90° with respect to the oscillation of the prong. Furthermore we have two signals: one proportional to the change in resonant frequency $\Delta f$ (relative to the free resonance frequency $f$) and the excitation voltage $V_{exc}$. The last signal is fed into a feedback loop controller (PID) which is responsible for adjusting the tip-sample distance to maintain the value of $V_{exc}$ constant at a reference value set by the user.

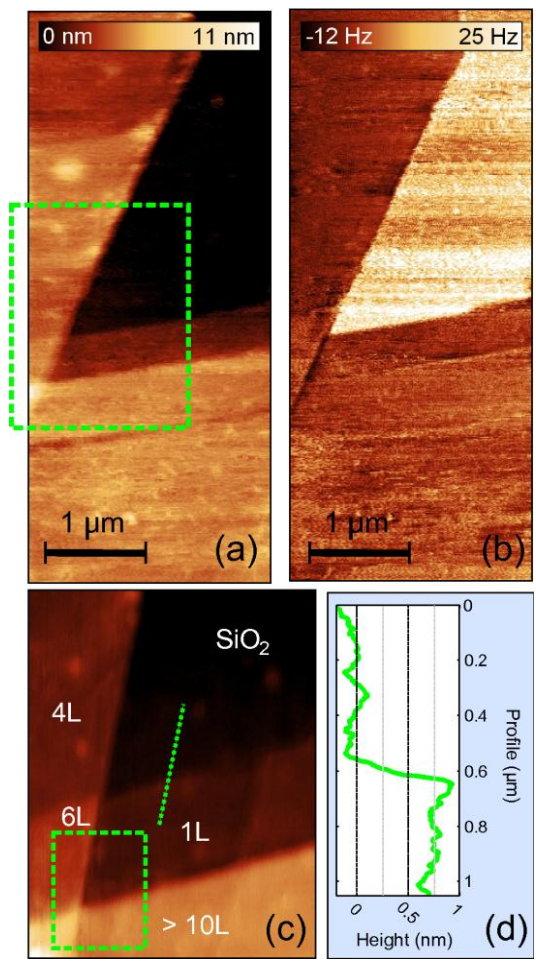

**Figure S5:** (a) FM-AFM topography of a flake containing a variable number of graphene layers deposited on top of a 285 nm SiO$_2$/Si surface. (b) Simultaneously measured frequency shift which can be used to discriminate the substrate from the flake regions. (c) A more detailed AFM topography of the region marked by a dashed square in (a). (d) Height profile along the dashed line in (c) showing a region with a thickness of $0.75 \pm 0.1$ nm compatible with the thickness of graphene measured with AFM.

The STM topographic images were acquired using the dynamic STM mode, keeping the time averaged tunneling current constant while the tip oscillates at the resonance frequency of the *qPlus* sensor[12] (~ 32.1 kHz), see Figure S6. The tip oscillation amplitude was 0.2 nm$_{RMS}$. Variations in the resonance frequency of the *qPlus* sensor were measured with the phase locked loop of a *Nanonis OC4 Oscillation Controller*. These resonance frequency shifts are related to the tip-sample force gradient.[12] The tunneling parameters were: a time averaged tunneling current of 0.8 nA and a bias voltage at the sample side of +100 mV.

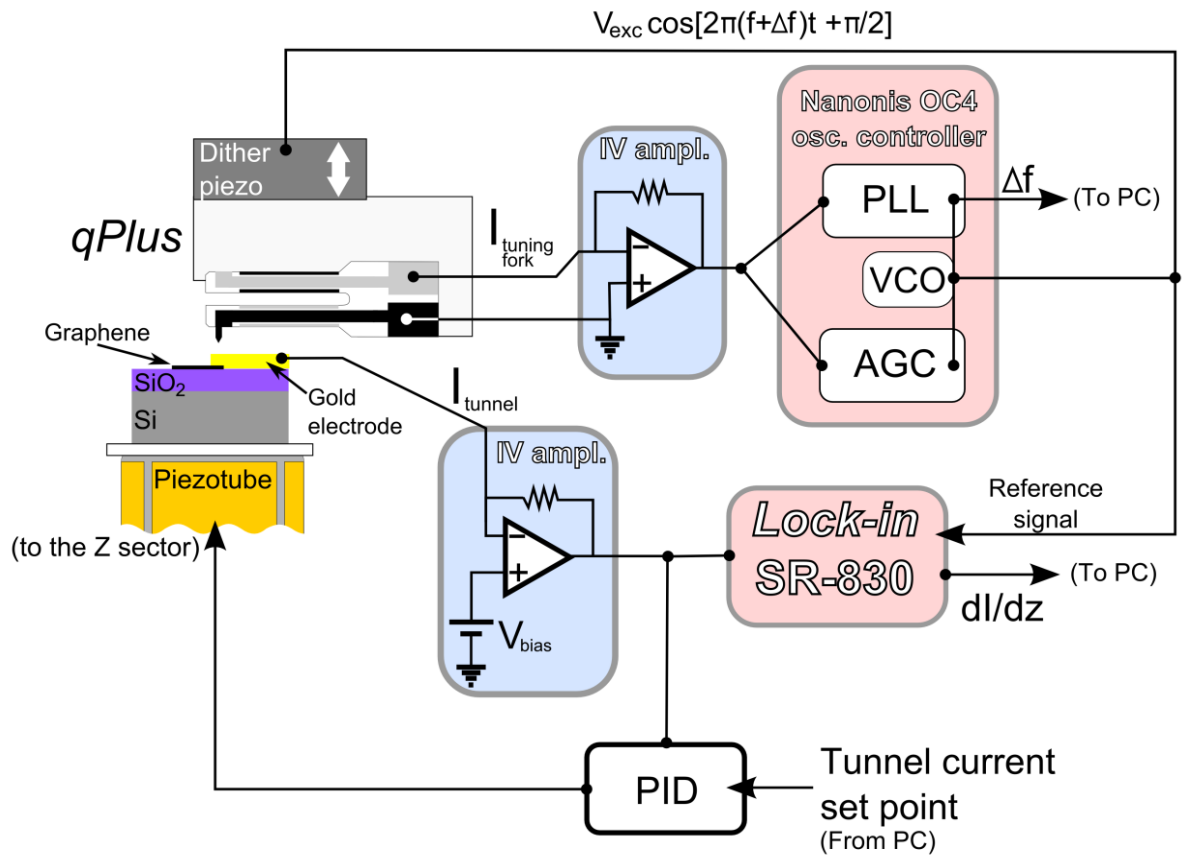

**Figure S6:** Schematic diagram of the experimental setup used in the combined STM/AFM measurements. A current-to-voltage converter (gain ×10$^9$) is connected to the graphene crystal through a gold electrode to measure the tunneling current ($I_t$) between tip and sample. The converter output is fed into a feedback control circuit (PID) that adjusts the tip-sample distance to maintain the value of the tunneling current $I_t$ constant at a reference value set by the user. Simultaneously with the STM mode operation, the *qPlus* sensor is mechanically excited with a piezoelectric actuator. As the resonance frequency of the *qPlus* is well above the time constant of the feedback control loop, it can not follow the oscillation of the tip and keeps constant the time averaged tunneling current during the oscillation.[13, 14] The piezoelectric current generated by the oscillation of the (free) prong of the *qPlus* sensor is preamplified by a current-to-voltage converter (gain ×10$^8$). The converter is connected to a *Nanonis OC4 Oscillation Controller* with a phase locked loop (PLL) and an automatic gain control (AGC). The output of *OC4 Nanonis* provides an excitation signal, which is connected to a dither piezo, at the resonance frequency of the sensor ($f + \Delta f$) and phase shifted 90° with respect to the oscillation of the prong. The change in resonant frequency $\Delta f$ (relative to the free resonance frequency $f$) is measured. This signal is proportional to the force gradient acting between tip and sample during the scan. In addition, a lock-in amplifier *SR-830 Stanford Research* is used to obtain the derivative of the tunneling current against the tip-sample distance ($dI/dz$), which can be related to the tunneling current decay constant $\beta$.

The tunneling current decay constant, $\beta$, has been obtained from the derivative of the tunneling current against the tip-sample distance ($dI/dz$), measured with a *lock in amplifier* SR830 of Stanford Research Systems: $\beta = -(dI/dz)/I$, with $I$ fixed to 0.8 nA during the scan due to the feedback control loop.

- 19 -

## *Calibration of the tunnelling decay constant measurement*

We construct the local tunnelling decay constant maps by dividing the (*dI /dz*) map by the tunneling current *I* measured during the scan (almost constant due to the feedback control loop). We have found very convenient to use quasi static current *vs.* distance traces to calibrate these maps and thus to obtain quantitative values of the local tunneling decay constant. First, we measure several tens of current *vs.* distance traces on a multilayer region (> 10 layers). In this region the current shows an exponential dependence with the tip-sample distance with a well-defined value of the tunneling decay constant $\beta \approx 5.3$ nm$^{-1}$. Figure S7a shows the average of 20 current *vs.* distance traces measured at different spots on a multilayer region (> 10 layers). Second, we measure the local tunneling decay constant in the dynamic STM mode (Figure S7c) which shows a nearly constant value. Then we calibrate the *β* map, measured by dynamic STM, in order to make its mean *β* value equal to 5.3 nm$^{-1}$.

If we repeat this experiment in a monolayer region we observe that the current *vs.* distance traces depend exponentially on the tip-sample distance but the decay constant shows a larger fluctuation in different regions of the monolayer *β* ~ (4.3-5.3) nm$^{-1}$ (Figure S7b). In principle this inhomogeneity of the local tunneling decay constant can be probed by making a map of current *vs.* distance traces. However, this measurement is much slower than the dynamic STM measurement and thus the effect of the thermal drift is more severe in this kind of measurement. If we measure the local tunneling decay constant map with the dynamic STM mode (Figure S7d), using the calibration constant previously determined, we can observe how the value *β* is locally inhomogeneous and indeed varies from 4.3 nm$^{-1}$ to 5.3 nm$^{-1}$.

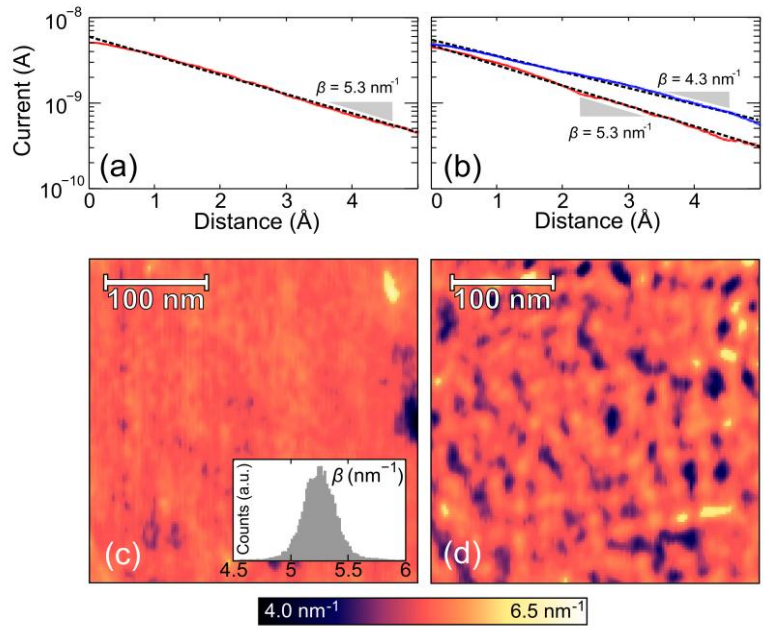

**Figure S7:** (a) Average of 20 current *vs*. distance traces measured at different spots on a multilayer region (>10 layers). (b) Two average current *vs*. distance traces measured at different sports on a graphene monolayer (with high and low tunneling decay constant). (c) Local tunneling decay constant map measured on the multilayer region, calibrated with the quasi static current *vs*. distance traces. The inset in (c) shows how the map presents a nearly constant value of the tunneling decay constant. (d) Local tunneling decay constant map measured on the monolayer region, using the calibration determined from the measurement on the multilayer region.

## *Cross-correlation between topography and β maps*

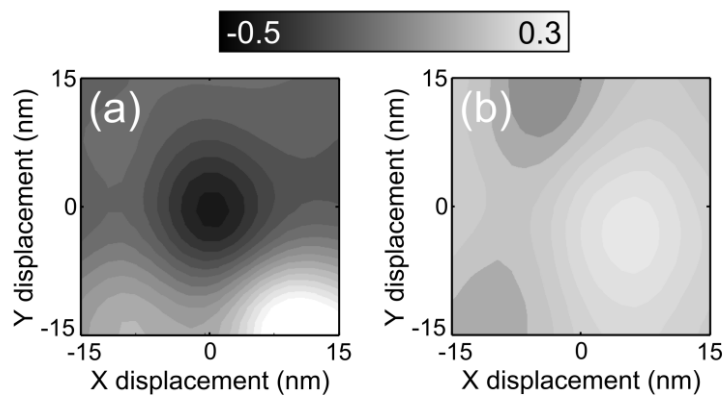

**Figure S8:** (a) Calculated normalized cross-correlation between the STM topography and the *β* map measured in the monolayer region shown in Figure S7(b). (b) Calculated normalized cross-correlation for the multilayer region shown in Figure S7(a). The well-defined anti-correlation obtained for the monolayer region indicates that the presence of the electronic inhomogeneities affects the STM topography. For the multilayer, the electric field generated by the charged impurities is screened and thus the topography is less affected by the presence of such impurities (weak correlation between the STM topography and the *β* map).

## *References for the supplemental material*

[1]    V. Geringer, D. Subramaniam, A. K. Michel, B. Szafranek, D. Schall, A. Georgi, T. Mashoff, D. Neumaier, M. Liebmann, M. Morgenstern, Appl. Phys. Lett. 2010, 96, 082114.

[2]    M. Ishigami, J. Chen, W. Cullen, M. Fuhrer, E. Williams, Nano Lett. 2007, 7, 1643.

[3]    J. Moser, A. Barreiro, A. Bachtold, Appl. Phys. Lett. 2007, 91, 163513.

[4]    Y. Dan, Y. Lu, N. J. Kybert, Z. Luo, A. T. C. Johnson, Nano Lett. 2009, 9, 1472.

[5]    N. Staley, H. Wang, C. Puls, J. Forster, T. N. Jackson, K. McCarthy, B. Clouser, Y. Liu, Appl. Phys. Lett. 2007, 90, 143518.

[6]    W. Bao, G. Liu, Z. Zhao, H. Zhang, D. Yan, A. Deshpande, B. LeRoy, C. N. Lau, Nano Research 2010, 3, 98.

[7]    Ç. Girit, A. Zettl, Appl. Phys. Lett. 2007, 91, 193512.

[8]    A. Castellanos-Gomez, N. Agrait, G. Rubio-Bollinger, Appl. Phys. Lett. 2010, 96, 213116.

[9]    P. Blake, E. W. Hill, A. H. C. Neto, K. S. Novoselov, D. Jiang, R. Yang, T. J. Booth, A. K. Geim, Appl. Phys. Lett. 2007, 91, 063124.

[10]   K. S. Novoselov, A. K. Geim, S. V. Morozov, D. Jiang, Y. Zhang, S. V. Dubonos, I. V. Grigorieva, A. A. Firsov, Science 2004, 306, 666.

[11]   7x precision zoom lenses from Edmund Optics. Reference 56-627.

[12]   F. J. Giessibl, Appl. Phys. Lett. 1998, 73, 3956.

[13]   K. Kobayashi, S. Kurokawa, A. Sakai, Jpn. J. Appl. Phys., Part 1 2004, 43, 4571.

[14]   Y. Yamada, A. Sinsarp, M. Sasaki, S. Yamamoto, Jpn. J. Appl. Phys., Part 1 2002, 41, 7501.